\documentclass[aps,pre,reprint,groupedaddress]{revtex4-2}
\usepackage[tbtags]{mathtools}
\usepackage{amsmath,amsfonts,mathdots,amssymb,yfonts,calc}
\usepackage{graphicx}
\usepackage{subfigure}

\begin{document}

%\preprint{}

\title{Charged Swarmalator in a Magnetic Adiabatic Trap}


\author{Vinesh Vijayan}
\email[]{vinesh.phy@gmail.com}
%\thanks{}
\altaffiliation{Department of Physics and Astronomy }
\affiliation{National Institute of technology, Rourkela, India-769008}
\author{Pranaya Pratik Das}
\email[]{519ph1005@nitrkl.ac.in}
%\thanks{}
\altaffiliation{Department of Physics and Astronomy }
\affiliation{National Institute of technology, Rourkela, India-769008}

%\author{Biplab Ganguli}
%\email[]{biplabg@nitrkl.ac.in}
%\thanks{}
%\altaffiliation{Department of Physics and Astronomy }
%\affiliation{National Institute of technology, Rourkela, India-769008}


\date{10/11/2021}

\begin{abstract}
In this letter, we provide a detailed numerical examination  of the dynamics of a charged swarmalator in an external magnetic field. We do so by adopting and then modifying the cyclically symmetric Thomas oscillator to  study the dynamics of a charged particle in  an external magnetic field. These dynamical behaviours for weak and strong field strength parameters fall under two categories; conservative and dissipative. For conservative regime the system shows a complex quasi-periodic attractor whose topology depends upon the initial condition at high field strengths. On decreasing the field strength parameter, there is a transition from adiabatic motion to chaos. In the dissipative regime, the system is chaotic for weak field strength and weak damping but shows a limit cycle for high field strengths. Such a behaviour mimics the dynamics of  magneto-bacterial microrobots.
\end{abstract}


%\maketitle must follow title, authors, abstract, and keyword
\maketitle

%\section{}
\section{\label{sc1}INTRODUCTION}
Swarmalators are particles that exhibit cooperative motion and synchronize their time behaviours represented by their internal phases\citep{Kevin_p,Kev1}. When coupled under suitable conditions, they can move spatially separated yet with synchronized time dynamics. As a primary requirement for swarming, Ishogin and his colleagues suggested a model where the state variables are cyclically symmetric and bidirectional coupling between spatial and phase dynamics\citep{Igo}. These particles can help model biological and physical systems whose self-assembly and synchronization interact. It also finds applications in microbotics\citep{Jso,Isl,Isl1,Shu,All} and in the study of magnetic particles in colloidal suspensions\citep{Wang,Abd,Ada}.\\
\\
One such model which is helpful to understand swarmalator dynamics is the cyclically symmetric model suggested by Rene Thomas. Physically the model represents a spherical particle of smaller size moving in a force field. The introduction of this model enables us to understand the role of feedback loops in generating chaotic behaviours\citep{th099}. The system shows cyclic symmetry of its state variables where its own phase dynamics decides its motion in space. When the state variables are the components of velocities, it can model active Brownian particles\citep{Speck16}. There is only one control parameter,  the dimensionless damping, that controls the effect of the environment. The root to  chaos and symbolic dynamics are very well-studied problems\citep{spo07}. Two coupled oscillators with linear and nonlinear coupling are also studied in \citep{Vin} and its synchronization properties are established. The pattern formation in connection with the active Brownian motion is also studied with nonlinear coupling scheme in\citep{vinesh1} while with a linear coupling and nonidentical oscillators in \citep{Vasi}. The cyclically symmetric Thomas system is given by Equation(\ref{E1})\\
\begin{equation}
  \begin{aligned}
  \frac{dX}{dT} &= -bX + sin(Y) \\
  \frac{dY}{dT} &= -bY + sin(Z) \\
  \frac{dZ}{dT} &= -bZ + sin(X)  
  \end{aligned}
  \label{E1}
\end{equation}
Where $b$ is the damping parameter. Exceptional properties materialise when a charged particle interacts with external magnetic field. The use of external magnetic field enables us to constrain and order the motion of the particle.  Whether it is a study related to astrophysical plasma or the swarm control of microbotics, the most straightforward situation will be to understand a single particle dynamics in a given external field. Studying a single charged swarmalator in a magnetic field is important as the subsequent dynamics could allow us to understand many such particles motion in cohesive groups. This paper considers a study on the dynamics of a charged Thomas oscillator in an external magnetic field.  With the magnetic field switched on, we have one more control parameter to control the dynamics and chaos in the system. We study the dynamics by  considering two cases, conservative($b=0$) and dissipative($b\neq 0$), for weak and strong field strength parameters. Unidirectional and isotropic constant magnetic fields have been used in this computational study.\\
\\
The paper is organized in the following way. In  (\ref{sc2}), we discuss the modelling of a charged swarmalator. Section (\ref{sc3}) comprises of an elaborate discussion on the dynamics of the oscillator for zero damping with unidirectional and isotropic magnetic fields. In (\ref{sc4}), We consider non-zero damping and study the dynamics  for both unidirectional and isotropic magnetic fields. Section (\ref{sc5}) is for our various results and discussion.
\section{\label{sc2}MODELING AND DIMENSIONAL ANALYSIS FOR A CHARGED SWARMALATOR IN AN EXTERNAL MAGNETIC FIELD}
Consider the motion of a charged particle in a fluid whose equations of motion are given below with the assumption that there is a constant magnetic field along the z-direction.  Here, $m$ is the mass of the particle, $\gamma$, the damping coefficient, $F$, is the strength of the interaction force field and the $v_{i}$'s, where $i=x,y,z$, are the components of velocities. The components of velocities are coupled with a cyclic symmetry.
\begin{align}
m\frac{dv_x}{dt}&= -\gamma v_x + F sin(\frac{v_y}{v_c}) + qv_yB_z\label{E2}\\
m\frac{dv_y}{dt}&= -\gamma v_y + F sin(\frac{v_z}{v_c}) - qv_xB_z\label{E3}\\
m\frac{dv_x}{dt}&= -\gamma v_z + F sin(\frac{v_x}{v_c})\label{E4} 
\end{align}
where $B_z$ is the magnetic field along the $z$- direction. Define dimensionless velocities and time as
\begin{eqnarray} 
X = \frac{v_x}{v_c}, Y =\frac{v_y}{v_c},Z=\frac{v_z}{v_c}, T=\frac{t}{\tau}
\label{E5}
\end{eqnarray}
where, $v_c$ is the characteristic speed and $\tau$ is the characteristic time such that the characteristic length scale can be defined to be $l=v_c \tau $. Putting Equation(\ref{E5}) in Equation(\ref{E2}) one will get the following.
\begin{equation}
\begin{split}
\frac{dX}{dT}& =-\frac{\gamma v_c \tau}{mv_c}X +\frac{F \tau}{mv_c}sin(Y) + \frac{qB_zv_c \tau}{mv_c} Y\\
             & =-\frac{\gamma \tau}{m}X +\frac{F\tau}{mv_c}sin(Y) + \frac{qB_z\tau}{m}Y 
\end{split}
 \label{E6}
\end{equation}
The dimensionless parameters are defined to be  $b=\frac{\gamma \tau}{m}$(damping parameter), $c=\frac{qB_z\tau}{m}$(field strength parameter) and set $\frac{F\tau}{mv_c}=1$. Following the similar procedure for Equations(\ref{E3},\ref{E4}) one will end up with a modified Thomas oscillator which is charged and placed in an external magnetic field along the z-direction.
\begin{equation}
  \begin{aligned}
\frac{dX}{dT}&=-bX + sin(Y) + cY\\
\frac{dY}{dT}&=-bY + sin(Z) - cX\\
\frac{dZ}{dT}&=-bZ + sin(X) 
  \end{aligned}
\label{E7}
\end{equation}
For Isotropic magnetic field we have
\begin{equation}
  \begin{aligned}
\frac{dX}{dT}&=-bX + sin(Y) + c(Y-Z)\\
\frac{dY}{dT}&=-bY + sin(Z) + c(Z-X)\\
\frac{dZ}{dT}&=-bZ + sin(X) + c(X-Y)
\end{aligned}
\label{E8}
\end{equation}
In a compact form one can write,
\begin{eqnarray}
{\bf{f}}({\bf{x}};b,c) = {\bf{g}}({\bf{x}};b) + {\bf{h}}({\bf{x}};c)
\label{E9}
\end{eqnarray}
where {\bf{x}} stands for the dimensionless velocities and b \& c are dimensionless parameters as defined above. {\bf{g}}({\bf{x}};b) is the Thomas system and {\bf{h}}({\bf{x}};c) is the Lorenz force term in the dimensionless form.
\section{\label{sc3}Dynamics For Zero Damping ({\lowercase{b}} = 0)}
\subsubsection{Unidirectional Magnetic Field}
\begin{figure}[hbt!]
    \centering
    \subfigure[]{\includegraphics[height=4cm,width=7cm]{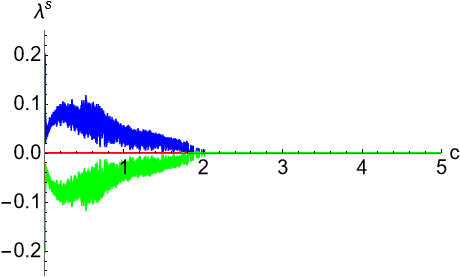}}\\ 
    \subfigure[]{\includegraphics[height=3.5cm,width=7cm]{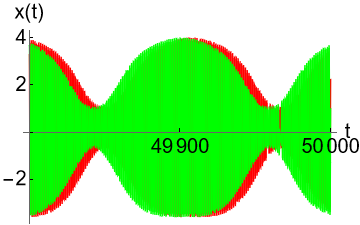}} 
   
    \caption{ (a) Lyapunov spectrum (b) Time series showing linearly diverging trajectories for two nearby initial conditions showing complex quasi periodic oscillations for $c=5$ }
    \label{LYA1}
\end{figure}

\begin{figure}[hbt!]
    \centering
    \subfigure[]{\includegraphics[height=3.4cm,width=4cm]{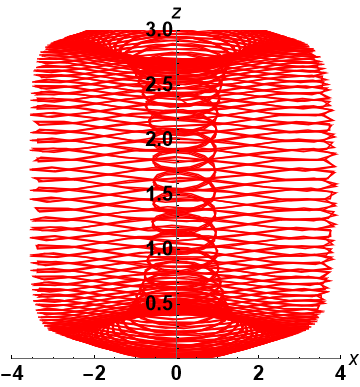}} 
    \subfigure[]{\includegraphics[height=3.4cm,width=4cm]{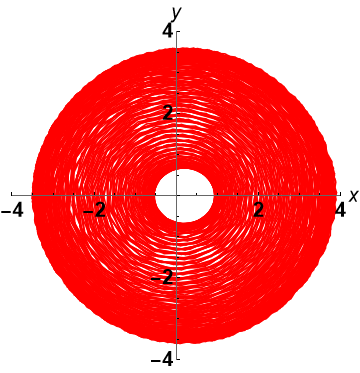}} 
    \subfigure[]{\includegraphics[height=3.4cm,width=4cm]{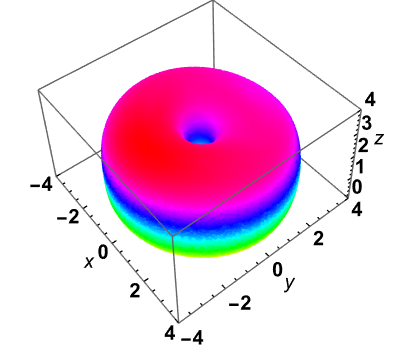}}
    \caption{Phase space plots for a charged Thomas oscillator under unidirectional magnetic field with $c=5$ (a)X-Z phase plot (b) X-Y phase plot (c) The complex quasi-periodic attractor}
    \label{Pha}
\end{figure}

In this section, the effect of a unidirectional magnetic field on a charged Thomas oscillator is considered under zero damping. The magnetic field direction is chosen to be along the z-direction, ${\bf B}= B{\bf {\hat{k} }}$. The dynamical equations, in this case, will be given by setting $b=0$ in Equation(\ref{E7}). The flow generated by this system of coupled equations are conservative since the divergence of flow is zero, ${\bf \nabla \cdot f}=0 $. Further, the sum of the Lyapunov exponents, in this case, is zero supporting this fact. With $b=0$ we have only one control parameter here, the field strength parameter$c$,  by tuning which one can explore the complex dynamics of the system. From  FIG(\ref{LYA1}-(a)) it is observed that the system behaviour is chaotic up to $c\approx 2$ for which the largest Lyapunov exponent(blue) is positive. After that, all the exponents are zero pointing towards complex quasi-periodic oscillations and are very close to regular dynamics FIG(\ref{LYA1}-(b)). The phase space of the system for $c=5$ is given in FIG(\ref{Pha}). \\
\\
Since the system is conservative the mechanism leading towards chaos is different from dissipative systems. In conservative systems, the phase volume must be conserved. Since the energy is conserved, the trajectory should be folded back in to the phase space. An important related phenomena is the phase-locking on to nonlinear resonance. In this phenomena, if the ratio of two frequencies associated with the dynamics approaches an integer  or otherwise a rational fraction nonlinearities causes phase locking of nearby trajectories on to this resonance\citep{Hand}. 
\\
A moving lump of charge is like a current flow that can generate a magnetic field around it.  We call this an induced magnetic field ${\bf B_i}$. The strength of the induced magnetic field is decided by the initial velocities, acceleration of the charge, amount of charge, and the strength of the external magnetic field. So for a fixed charge and mass, the only way one can control the induced magnetic field is by choice of initial conditions and by varying the external field. Due to the superposition of the two magnetic fields, the applied magnetic field and the built-in magnetic field, we have a coaxial double- helical motion, with the inner helix diverging out into the outer helix and the vice-versa(the outer helix converging in into the inner helix). The conversion and diversion happens at opposite poles due to drift motion and they can switch polarity depending on the nature of charge and/or the direction of magnetic field. Also because this drift motion is an accelerated motion radiations will be
emitted at both ends of these helix’s resembling a magnetic wormhole. One can say the particle is trapped in an external magnetic field. \\
\begin{figure}[hbt!]
    \centering
    \subfigure[]{\includegraphics[height=3.5cm,width=4cm]{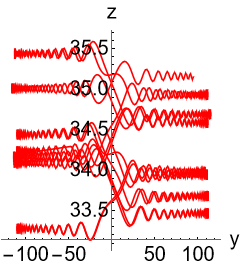}}     
    \subfigure[]{\includegraphics[height=3.5cm,width=4cm]{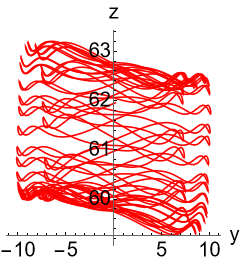}} \\    
    \subfigure[]{\includegraphics[height=3.5cm,width=4cm]{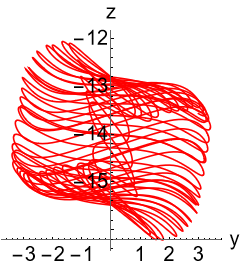}}     
    \subfigure[]{\includegraphics[height=3.5cm,width=4cm]{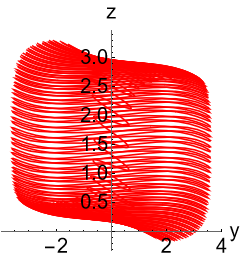}}

    \caption{The figure shows the projections of attractors on to $y-z$ plane for magnetic field strength parameter (a) c = 0.3(Chaotic Motion) (b) c = 1.2(Chaotic Motion) (c) c = 1.3(Chaotic Motion) (d) c = 3.5 (Complex Quasi-periodic Motion)respectively for constant unidirectional  external magnetic field .}
    \label{Uni_PH}
\end{figure}

\begin{figure}[hbt!]
    \centering
    
    \subfigure[]{\includegraphics[height=3.5cm,width=4cm]{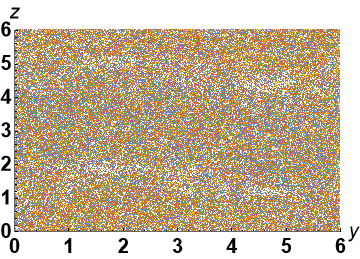}}     
    \subfigure[]{\includegraphics[height=3.5cm,width=4cm]{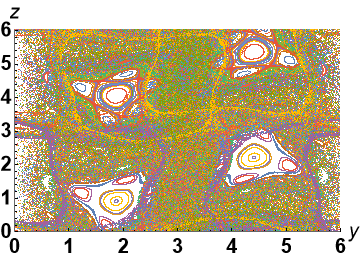}}\\    
    \subfigure[]{\includegraphics[height=3.5cm,width=4cm]{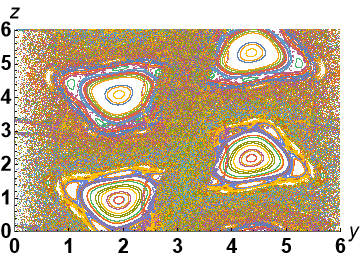}}    
    \subfigure[]{\includegraphics[height=3.5cm,width=4cm]{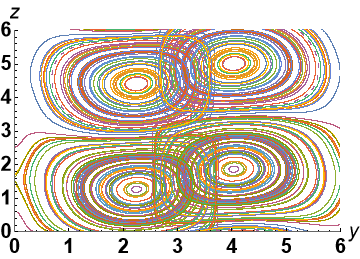}} 
   
    \caption{The figure shows the bifurcation taking place in the system via Poincar\'e - sections for magnetic field strength parameter (a) c = 0.3(Chaotic Motion) (b) c = 1.2(Chaotic Motion) (c) c = 1.3(Chaotic Motion) (d) c = 3.5(Complex Quasi-periodic Motion) respectively for constant unidirectional external magnetic field .}
    \label{Uni_PR}
\end{figure}

\begin{figure}[hbt!]
    \centering
   
    \subfigure[]{\includegraphics[height=3.5cm,width=4cm]{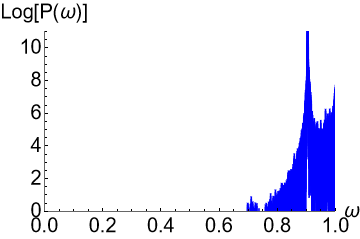}}   
    \subfigure[]{\includegraphics[height=3.5cm,width=4cm]{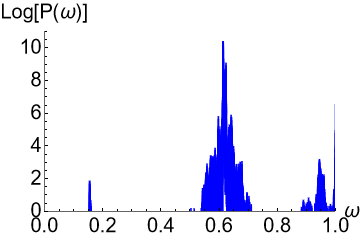}}\\   
    \subfigure[]{\includegraphics[height=3.5cm,width=4cm]{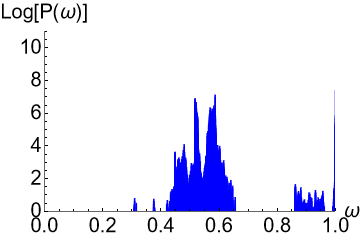}}     
    \subfigure[]{\includegraphics[height=3.5cm,width=4cm]{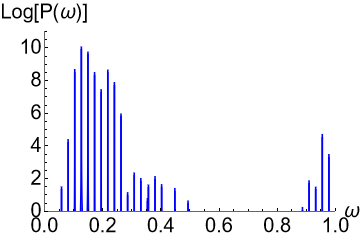}}
   
    \caption{The figure shows  the power spectra for magnetic field strength parameter (a) c = 0.3(Chaotic Motion) (b) c = 1.2(Chaotic Motion) (c) c = 1.3(Chaotic Motion) (d) c = 3.5(Complex Quasi-periodic Motion) respectively for constant unidirectional external magnetic field .}
    \label{Uni_PW}
\end{figure}

\noindent According to classical electromagnetic theory, if a particle with charge $q$ and mass $m$ moving with a velocity ${\bf v}$  is put into an external magnetic field that will be acted on by a Lorenz force perpendicular to both the magnetic field vector and  velocity vector. In a strong uniform magnetic field,  the motion of the charged particle is such that they gyrate around the direction of the external magnetic field. The frequency of the gyrated motion will be given by
\begin{equation}
\omega_c = \frac{q{\bf B}}{m}=\frac{2 \pi r_c }{\mathcal{T}}
\label{E11}
\end{equation}
and is also known as the cyclotron frequency, where $r_c$ is the cyclotron radius and $\mathcal{T}$ the period of cyclotron motion. The frequency of gyration depends on the strength of the magnetic field {\bf B}, charge and mass of the particle. In the absence of other external forces, in a homogeneous magnetic field, the motion of the charged particle will be a combination of gyration as well as uniform rectilinear motion which will result in a helical trajectory with a given frequency along the field lines. The radius of gyration will be smaller wherever the field strength is a maximum and larger wherever the field strength is a minimum\citep{Frank}. \\
\\
Whenever the particle moves through an in-homogeneous magnetic field, a force will start to act at right angles to the field giving the particle a direction of motion perpendicular to both the force and the magnetic field. The in-homogeneity in the magnetic field can be parallel or perpendicular to the direction of the applied magnetic field.  Perpendicular in-homogeneity means there are compression's and rarefaction's of field lines. The variations in the field parallel to the applied field direction lead to curved field lines. The former leads to gradient drift and the latter to the centrifugal drift. Assume {\bf R} is the radius of curvature of the magnetic field lines then the magnitudes and directions of these two drifts are given by
\begin{eqnarray}
{\bf v}_G & = \frac{-mV_{\perp}^2}{2qB^2}({\bf B \times \nabla B})\\
{\bf v}_C & = \frac{-mV_{\parallel}^2}{q B^2 R^2}({\bf R \times B})
\label{E12}
\end{eqnarray}
The minus sign here indicates that the motion of the particle will be opposite for positive and negative charges. One can distinguish between the drift motion and cyclotron motion in the presence of a strong applied magnetic field.  In the presence of a strong magnetic field, the distortion of cyclotron orbits due to in-homogeneities or other forces will be small. The changes in period and the radius of gyration are then small compared to the characteristic time scale and length scales that characterize all other relevant quantities that characterize these orbits. The motion that satisfies all these conditions are called adiabatic motion ie,  $\omega_c >>  \frac{1}{\tau}, r_c << l$.\\
\\
When the particle motion satisfies the adiabatic conditions, then it can be treated as a combination of three mutually exclusive motions. The rectilinear motion along the field, the gyrated motion around the field, and the drift motion in a direction perpendicular to the field. Suppose the condition for adiabatic motion is not satisfied then there will be confusion among the latter two motions and will lead to complex dynamics and chaos.  In this case,  for the dynamics of the charged swarmalator, the adiabatic conditions are not satisfied in the limit of the weak applied magnetic field.  As we decrease the magnetic field strength parameter, the system falls onto nonlinear resonance and from there to chaos. In this system, when the strength of the field strength parameter is decreased, the strength of the nonlinear resonance will increase and they will overlap. As a result of this it will never follow a smooth regular path in the phase space but wander chaotically. The smooth curves on the invariant-tori breaks because of the competition between these nonlinear resonances.\\
\\
Projections of the chaotic attractors for different values of $c$ are given in FIG(\ref{Uni_PH}). The corresponding bifurcation of the system with field strength parameter is shown via the Poincar\'e sections. As we see from FIG(\ref{Uni_PR}-(a)), for the weak magnetic field strength parameter we have a chaotic sea. For intermediate values of the parameter, we can see mixing and for higher values, the trajectories are nearing regular trajectories. From FIG(\ref{Uni_PR}-(b,c)), one can observe quasi periodic orbits surrounded by a KAM surface. From the Lyapunov spectra(FIG(\ref{LYA1})), the  Lyapunov exponents are  zero for higher values of the parameter showing that there is complex quasi-periodic oscillations, very near to regular motion. The resonance overlap criteria make use of the fact when distinct resonances overlap, no KAM tori can exist between them. The strength of the external field at which this occurs marks the commencement of chaos. The power spectrum of the system for the same parameter values are also given in FIG(\ref{Uni_PW}). The chaotic nature of the system is shown by a wide frequency distribution whereas for regular systems, the power spectrum consists of single lines at harmonics and sub harmonics. \\

\subsubsection{Isotropic Magnetic Field} 

The dynamics for an isotropic external magnetic field is given by setting $b=0$ in Equation(\ref{E8}). This case also shows conservative dynamics and at higher values of the field strength parameter, the system reaches to the complex quasi periodic attractor much earlier than that in the previous case, around $c\approx0.97$. Most importantly, the attractor gets aligned with the resultant field direction. The Laypunov spectrum is given below in FIG(\ref{LYA2}).\\
\begin{figure}[hbt!]
    \centering
    \subfigure[]{\includegraphics[height=4cm,width=7cm]{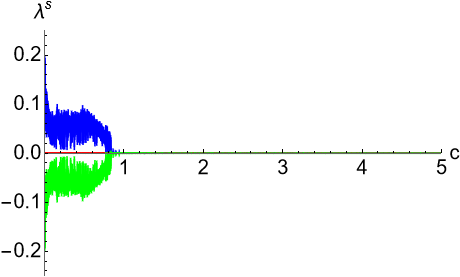}}\\ 
    \subfigure[]{\includegraphics[height=3.5cm,width=7cm]{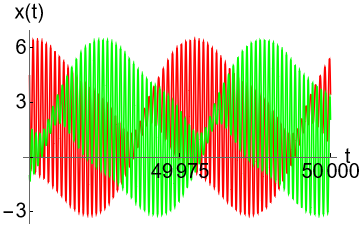}} 
   
    \caption{ (a) Lyapunov spectrum (b) Time series showing linearly diverging trajectories for two nearby initial conditions showing complex quasi periodic oscillations for $c=5$ }
    \label{LYA2}
\end{figure}
\begin{figure}[hbt!]
    \centering
    \subfigure[]{\includegraphics[height=3.5cm,width=4cm]{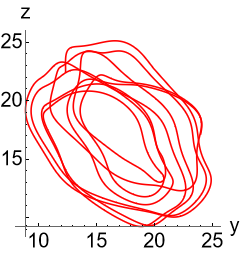}}     
    \subfigure[]{\includegraphics[height=3.5cm,width=4cm]{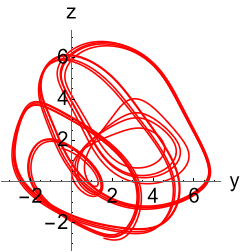}} \\   
    \subfigure[]{\includegraphics[height=3.5cm,width=4cm]{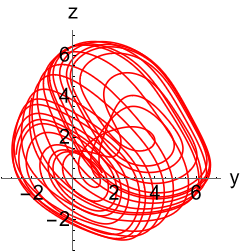}}   
    \subfigure[]{\includegraphics[height=3.5cm,width=4cm]{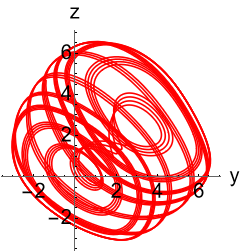}} 
   
    \caption{The figure shows the projections of attractors on to $y-z$ plane for magnetic field strength parameter (a) c = 0.3(Chaotic Motion) (b) c = 0.7(Chaotic Motion) (c) c = 0.9(Complex Quasi-periodic Motion) (d) c = 1.5(Complex Quasi-periodic Motion) respectively for constant isotropic external magnetic field.}
    \label{Iso_PH}
\end{figure}

\begin{figure}[hbt!]
    \centering    
    \subfigure[]{\includegraphics[height=3.5cm,width=4cm]{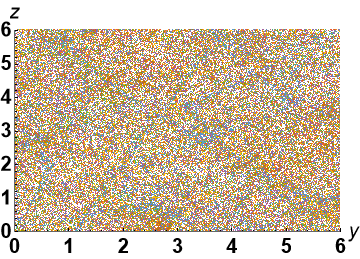}}    
    \subfigure[]{\includegraphics[height=3.5cm,width=4cm]{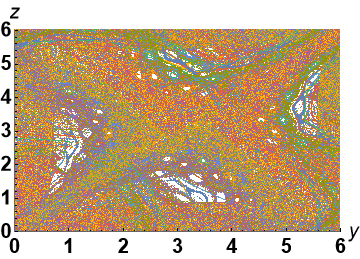}}\\    
    \subfigure[]{\includegraphics[height=3.5cm,width=4cm]{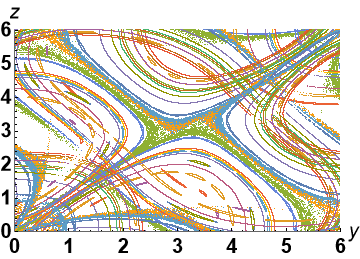}}   
    \subfigure[]{\includegraphics[height=3.5cm,width=4cm]{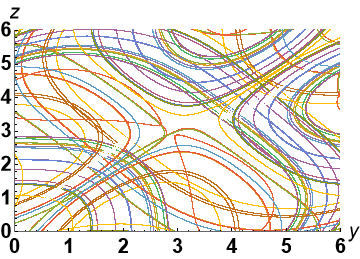}}

    \caption{The figure shows the bifurcation taking place in the system via Poincar\'e - sections  for magnetic field strength parameter (a) c = 0.3(Chaotic Motion) (b) c = 0.7(Chaotic Motion) (c) c = 0.9(Complex Quasi-periodic Motion) (d) c = 1.5(Complex Quasi-periodic Motion) respectively for constant isotropic external magnetic field.}
    \label{Iso_PR}
\end{figure}

\begin{figure}[hbt!]
    \centering   
    \subfigure[]{\includegraphics[height=3.5cm,width=4cm]{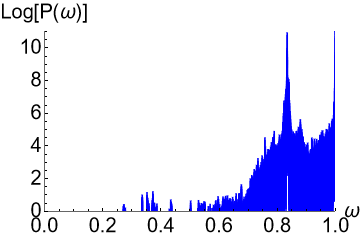}}   
    \subfigure[]{\includegraphics[height=3.5cm,width=4cm]{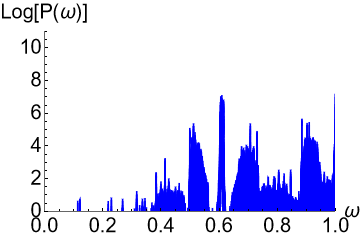}}\\    
    \subfigure[]{\includegraphics[height=3.5cm,width=4cm]{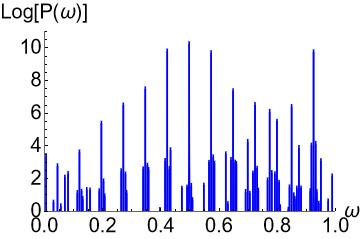}}    
    \subfigure[]{\includegraphics[height=3.5cm,width=4cm]{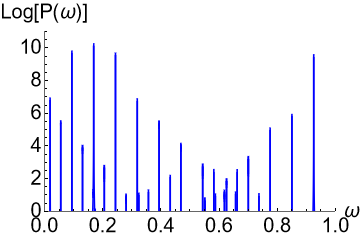}}
   
    \caption{The figure shows the  power spectra for magnetic field strength parameter (a) c = 0.3(Chaotic Motion) (b) c = 0.7(Chaotic Motion) (c) c = 0.9(Complex Quasi-periodic Motion) (d) c = 1.5(Complex Quasi-periodic Motion) respectively for constant isotropic external magnetic field.}
    \label{Iso_PW}
\end{figure}
\noindent The projections of attractors are shown in FIG(\ref{Iso_PH}) and the bifurcation diagram via Poincar\'e sections  are in FIG(\ref{Iso_PR}). The corresponding power spectrum is given in Figure(\ref{Iso_PW}).
\section{\label{sc4}Dynamics Under Dissipation ({\lowercase{b}} $ \neq $ 0)}
The second part of the paper deals with the dissipative case of the modified Thomas oscillator in an external magnetic field. This is motivated by the fact that when a colloidal particle is allowed to move in solution with certain viscosity, they move around in that space in such a way that it will create a feedback loop between phase and space dynamics. If such a particle possesses magnetic properties and if it is placed in an external magnetic field they start to oscillate and establish coupling between nearby particles and they tend to synchronize their behaviour. So we propose that the modified Thomas oscillator with the nonzero damping parameter can model such a particle, and this may lead to the understanding of dynamics of a collection of such particles.\\
\subsubsection{Unidirectional Magnetic Field}
For this study, first of all, we consider the unidirectional applied magnetic field as modelled by Equation(\ref{E7}). Two different values of $b$ are chosen; a low value$(0.05)$ and a high value$(0.18$). The bifurcation diagrams and Lyapunov spectra are given below in FIG(\ref{L_B_05Z},\ref{L_B_18Z}).\\
\begin{figure}[hbt!]
    \centering   
    \subfigure[]{\includegraphics[height=4cm,width=7cm]{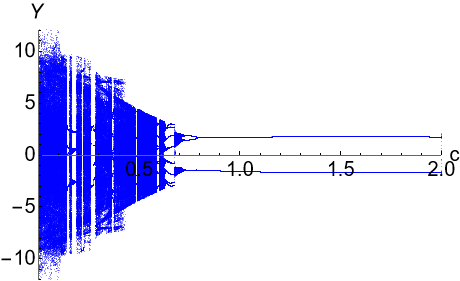}}\\ 
    \subfigure[]{\includegraphics[height=4cm,width=7cm]{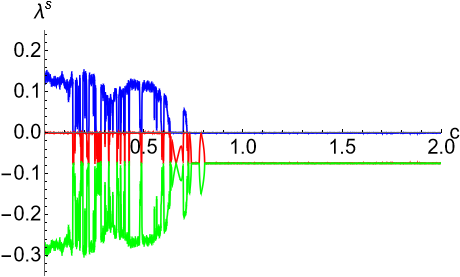}}    
      
    \caption{The figure shows the(a) bifurcation diagram and (b) Lyapunov spectrum for weak damping($b=0.05$) of  a charged swarmalator in a fluid with unidirectional magnetic field.}
    \label{L_B_05Z}
\end{figure}

\noindent For weak damping$(b=0.05)$, the dynamics of the oscillator is such that for higher values of the field strength parameter there is a limit cycle in the range $c=0.8 - c=2$. The Largest Lyapunov exponent is zero throughout this range(blue). Around $c=0.8$, the system undergoes a period-doubling cascade and becomes chaotic for further decrease in the field strength parameter. One can observe many quasi-periodic windows in the range $c\approx0.13 - 0.65$. In these periodic windows, one can observe transient chaos\citep{Tel1}. Finally in the range $c=0-c=0.13$ there is well-established chaos. For higher values of damping coefficients, $(b=0.18)$, The chaotic motion gets confined to a small window of parameter $c$. From a larger value of $c=1.67$, the system shows a fixed point. But at $c=1.67$, the system undergoes Hopf-bifurcation giving birth to a limit cycle. From $c=0.57 - 1.67$, we have limit cycle oscillations. Thereafter, the system undergoes windows of chaotic motion embedded in complex quasi-periodic oscillations.\\
\begin{figure}[hbt!]
    \centering   
    \subfigure[]{\includegraphics[height=4cm,width=7cm]{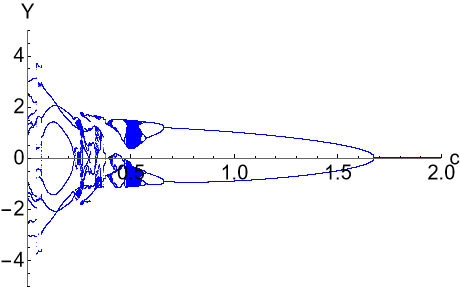}}\\    
    \subfigure[]{\includegraphics[height=4cm,width=7cm]{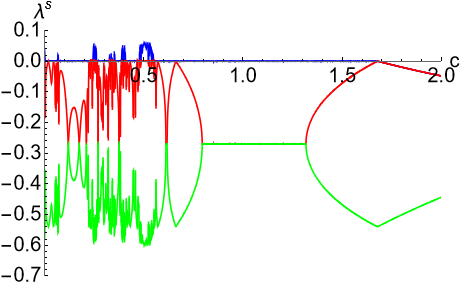}}
   
    \caption{The figure shows the(a) bifurcation diagram and (b) Lyapunov spetrum for high damping($b=0.18$) of  a charged swarmalator in a fluid with unidirectional magnetic field.}
    \label{L_B_18Z}
\end{figure}

\begin{figure}[hbt!]
    \centering   
    \subfigure[]{\includegraphics[height=3.5cm,width=4cm]{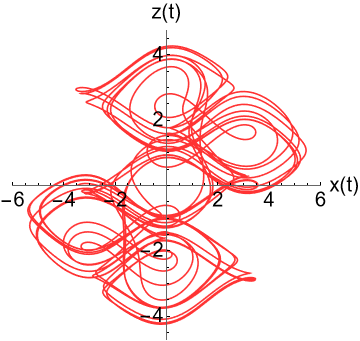}}   
    \subfigure[]{\includegraphics[height=3.5cm,width=4cm]{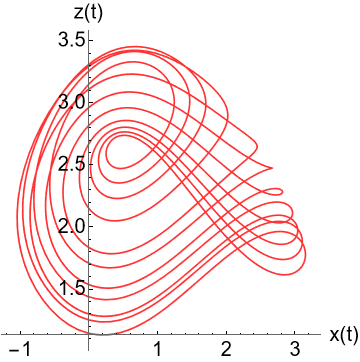}}
        
     \caption{The figure shows the projections of chaotic attractor(a) for $b=0.05$ and (b) for $b=0.18$ of  a charged swarmalator in a fluid with unidirectional magnetic field strength $c=0.53$.}
 \label{Uni_z_A}
\end{figure}

FIG(\ref{Uni_z_A}) shows the comparison of chaotic attractors for $b=0.05$ and $b=0.18$ for a fixed value of  the field strength parameter $c=0.53$.
\subsubsection{Isotropic Magnetic Field}
\begin{figure}[hbt!]
    \centering   
    \subfigure[]{\includegraphics[height=4cm,width=7cm]{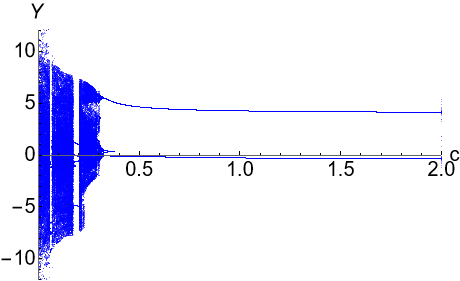}}\\ 
    \subfigure[]{\includegraphics[height=4cm,width=7cm]{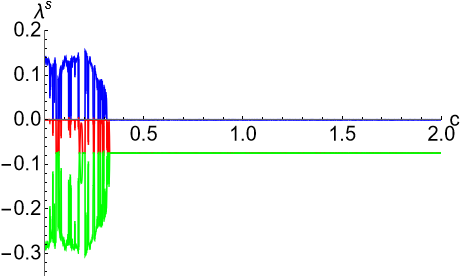}}    
      
    \caption{The figure shows the(a) bifurcation diagram and (b) Lyapunov spectrum for weak damping($b=0.05$) of  a charged swarmalator in a fluid with isotropic magnetic filed.}
    \label{L_B_05A}
\end{figure}

 Now we look at Equation(\ref{E8}) for the isotropic magnetic field applied to a charged Thomas system. For the weak value of damping, we have the following scenario that the oscillations prevail in the system for higher values of parameter $c$ for a damping $b=0.05$.  The system jumps to a period-doubling cascade towards chaos for a $c$ value around $0.35$ where it undergoes a Hopf bifurcation. One can observe complex quasi-periodic windows confined to a small parameter range in the chaotic sea. The largest Lyapunov exponent is zero(blue) from a higher value to $c=0.35$ showing the existence of a limit cycle in this range and is confirmed by the bifurcation diagram as shown in FIG(\ref{L_B_05A}(b)).\\

\begin{figure}[hbt!]
    \centering   
    \subfigure[]{\includegraphics[height=4cm,width=7cm]{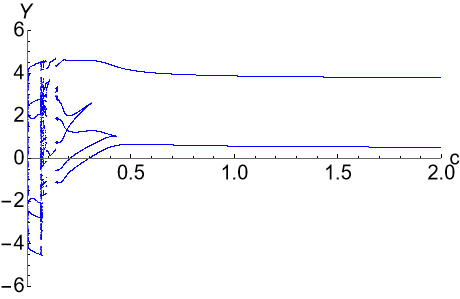}}\\ 
    \subfigure[]{\includegraphics[height=4cm,width=7cm]{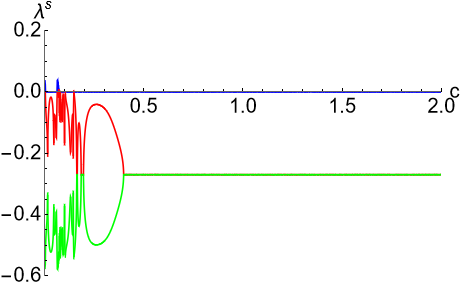}}    
      
    \caption{The figure shows the(a) bifurcation diagram and (b) Lyapunov spectrum for high damping($b=0.18$) of  a charged swarmalator in a fluid with isotropic magnetic filed.}
    \label{L_B_18A}
\end{figure}

\noindent   The scenario is  interesting when the damping coefficient is increased to $b=0.18$. Now, in this case, mostly we have oscillations except for a narrow window of the parameter value $c$ where it is chaotic. Chaoticity of the system was found to be very near to zero and for $c\approx0.067$. Otherwise for the whole range of the parameter values the system showed oscillations. Similarly as compared to the unidirectional field with high damping one can see that here in this case we are not having any fixed points. The Lyapunov spectrum and bifurcation diagram are given in FIG(\ref{L_B_18A}).

\begin{figure}[hbt!]
    \centering   
    \subfigure[]{\includegraphics[height=3.5cm,width=4cm]{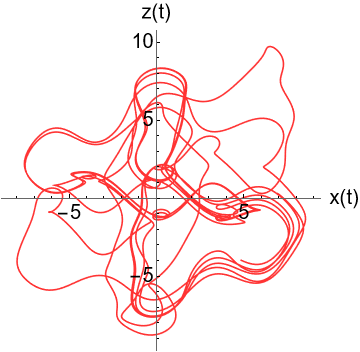}}   
    \subfigure[]{\includegraphics[height=3.5cm,width=4cm]{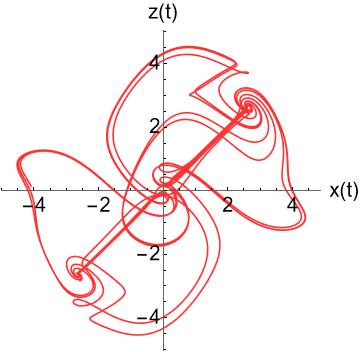}}
   
    \caption{The figure shows the projections of chaotic attractors(a) for $b=0.05$ and (b) for $b=0.18$ of  a charged swarmalator in a fluid with unidirectional magnetic field strength $c=0.069$.}
    \label{Iso_A_A}
\end{figure}

A comparison of chaotic attractors for $b=0.05$ and $b=0.18$ for a fixed value of  the field strength parameter $c=0.069$ is shown in FIG(\ref{Iso_A_A}).
\section{\label{sc5}Results and Discussions}
First of all, in the special case of conserved dynamics of the charged swarmalator, one can see that the volumes are preserved locally.  The transport of charge from one point to another point so that it is trying to keep some properties of the phase space a constant. The local curving of the trajectories in the phase space is such that the stretching and folding of the trajectories keep the phase volume conserved. Secondly, the motion of the charge is such that in the presence of an external magnetic field the kinetic energy of the charge is converted into internal magnetic energy and then this energy is liberated in the form of particle acceleration and radiation. The third important point is that an external magnetic field tends to orient a current loop such that the field of the circuit points in the direction of the applied magnetic field. Finally, the topology of the quasi periodic attractor can be changed by the choice of the initial condition, for example, from a straight line to a torus for the isotropic external magnetic field. The above-mentioned properties are important to understand the cause of events involved in plasma physics and astrophysics. Cosmic magnetic structures such as terrestrial magnetic storms, solar flare, and radio pulsar wind, etc are some examples where the dynamics of charged swarmalators can be used.\\
\\
In this study, for the conservative dynamics of the charged swarmalator in an external magnetic field, we find that the particle is trapped with quasi-periodic oscillations at high field strength parameter values. The external magnetic field confine a charged particle within a bounded domain of space. In this situation, there exists slowly varying dynamical quantities, the so-called adiabatic in-variants, which leads the system dynamics very close to the regular motion. When there is a destruction of the adiabatic in-variants the dynamics becomes chaotic. So one can say there is a particle confinement  through a magnetic adiabatic trap. By the adiabatic trap one can prevent the particle loss along the magnetic field direction. Such magnetic traps are corner stones of modern ultra-cold physics, reactor physics, quantum information processing, quantum optics, and quantum metrology. \\
\\
Modified Thomas system in the dissipative regime is good candidate to understand  magneto-bacterial micro-robotics.  Since the particle size is small, one can use a swarm of such particles for the  targeted delivery of drugs to cells. They are effortless to detect and navigate since they emit radiation. Soft micro-robots are particularly interesting  because by applying an external DC magnetic field, its trajectories can be controlled. Particularly interesting micro-robots are Xenobots which don't use any metal or semiconductors. Instead, they use biological tissues, and the advantage is that environmental pollution can be avoided as compared to traditional micro-robots. They can consume energy from the surroundings and can perform useful work, and these devices are biodegradable and bio-compatible. In all these cases the control is achieved by the control of the feedback mechanism. So the modified Thomas system will be a good candidate for understanding the dynamics of such micro-robots.
\section{\label{sc6}Conclusion}
In this paper, we have considered a charged swarmalator assuming that it is placed in a constant external magnetic field. The charged swarmalator is modelled by modifying the Thomas oscillator and dynamics are studied by considering two regimes of application; the conservative, and dissipative for weak and strong magnetic fields. In all the cases, the interplay between dynamics, external field, and the induced field leads to interesting properties. In the conservative regime, an adiabatic particle trapper is realized. The transition from adiabatic motion to chaos is through the destruction of adiabatic in-variance. In the dissipative regime, chaos can be controlled by the application of the external field. The transition to chaos in this case is via period-doubling cascades and the oscillatory nature for a wide range parameter values are suitable for the study of dynamics in microbotics.

\section*{Acknowledgements}
The first author greatly appreciates the discussions with Mr. Pranaya Pratik Das and his contributions towards finalizing the manuscript. Wish to acknowledge the support and motivation came from Mr.Ralu Johny and Mr.Ratheesh S Chandran.
\bibliography{Bibliography}

\end{document}